# Novel Scintillating Materials Based on Phenyl-Polysiloxane for Neutron Detection and Monitoring

M. Degerlier[1], S. Carturan[2,4], F. Gramegna[2], T. Marchi[2], M. Dalla Palma[2,3], M. Cinausero[2], G. Maggioni[2,4], A. Quaranta[2,3], G. Collazuol[4], J. Bermudez[2]

[1] University of Nevsehir, Science and Art Faculty Physics Department, Nevsehir Turkey.

[2] INFN - Laboratori Nazionali di Legnaro, Italy.

[3] University of Trento, Department of Industrial Engineering, Trento Italy.

[4] University of Padova, Department of Physics and Astronomy, Padova Italy.

**Abstract**

Neutron detectors are extensively used at many nuclear research facilities across Europe. Their application range covers many topics in basic and applied nuclear research: in nuclear structure and reaction dynamics (reaction reconstruction and decay studies); in nuclear astrophysics (neutron emission probabilities); in nuclear technology (nuclear data measurements and in-core/off-core monitors); in nuclear medicine (radiation monitors, dosimeters); in materials science (neutron imaging techniques); in homeland security applications (fissile materials investigation and cargo inspection). Liquid scintillators, widely used at present, have however some drawbacks given by toxicity, flammability, volatility and sensitivity to oxygen that limit their duration and quality. Even plastic scintillators are not satisfactory because they have low radiation hardness and low thermal stability. Moreover organic solvents may affect their optical properties due to crazing. In order to overcome these problems, phenyl-polysiloxane based scintillators have been recently developed at Legnaro National Laboratory. This new solution showed very good chemical and thermal stability and high radiation hardness. The results on the different samples performance will be presented, paying special attention to a characterization comparison between synthesized phenyl containing polysiloxane resins where a Pt catalyst has been used and a scintillating material obtained by condensation reaction, where tin based compounds are used as catalysts. Different structural arrangements as a result of different substituents on the main chain have been investigated by High Resolution X-Ray Diffraction, while the effect of improved optical transmittance on the scintillation yield has been elucidated by a combination of excitation/fluorescence measurements and scintillation yield under exposure to alpha and γ-rays.

## Introduction

A renewed interest in neutron detection has grown up not only in nuclear physics: neutron monitors are requested in nuclear power plants, spallation neutron sources, homeland security, nuclear medicine and materials analysis [1,2,3].

Organic scintillators, both liquid and solid, are an optimal choice as neutron detectors due to their low Z composition and high concentration of hydrogen atoms [4]. The most common liquids are based on xylene and pseudocumene solvents with the addition of different dyes and suitable waveshifters, so that emission is shifted to the blue region where the most standard photomultiplier tubes can work [5]. Liquid scintillators, suitable for large volumes and pulse shape analyses, display high quantum efficiency and fast response, though they are highly toxic and flammable [4]. Very recently, linear alkylbenzene (LAB) has been used as aromatic solvent for new liquid scintillators, based on its lower toxicity and volatility, though the issues of high flammability and waste disposal are still critical [6].

On the other hand, plastic scintillators, based on polyvinyltoluene (PVT), though displaying high quantum efficiency, are characterized by low radiation hardness and unfitness for pulse shape discrimination.

PSS-based scintillators have been object of deep investigation, owing to their intrinsic superior chemical and physical properties such as thermal and radiation resistance [7-9].

The aim of the present study is to review recently obtained results in this field and present newly developed formulation of polysiloxanes, which lead to further enhancement in the overall performances of the scintillators, preserving the demonstrated capability to detect both fast and thermal neutrons.

## Materials and Methods

In general, PSS can be vulcanized by addition polymerization, where a vinyl terminated resin reacts with a Si-H containing resin in presence of a Pt based catalyst. As a result of complete cross-linking, the structure is robust, non-sticky though elastic, able to withstand severe changes of temperature. Previous works on PSS formulations and cross-linked by Pt catalyzed addition report in detail the synthetic procedures [7-9].

Still another pathway to achieve vulcanized siloxanes consists in a condensation process, where a base siloxane resin, with terminal hydroxyl groups -OH reacts with a small quantity of cross-linker polydiethoxysilane. The two components are mixed in the weight ratio 100:10, in presence of a tin or titanium based catalyst (in this case dibutyltindilaurate DBTL was adopted, 1.3% wt.), as depicted in Fig. 1a, whereas in Fig. 1b a sketch of the structural arrangement after vulcanization has been figured out.

The use of Sn based catalyst instead of Pt can improve the scintillation efficiency, since Pt can interact with dye molecules, with negative consequences on their optical properties [10] and can easily form Pt nanoclusters, which absorb light in the blue region and, therefore, decrease the light yield of the scintillator.

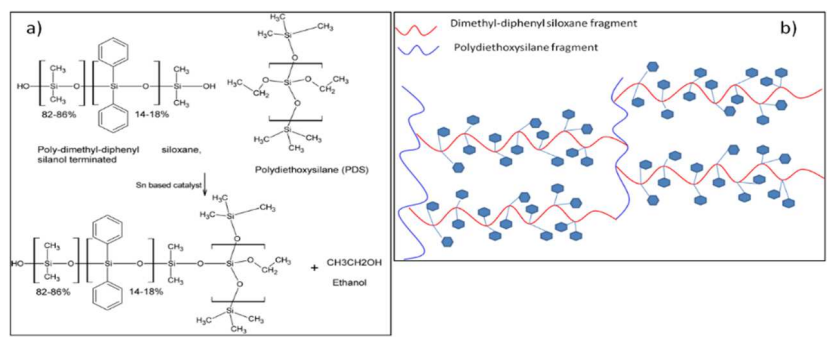

Fig. 1. a) reaction scheme and components used for the preparation by Sn-catalyzed condensation of the PSS scintillators and b) schematic drawing of the cross-linked siloxane.

The condensation reaction was carried out after addition to the silanol terminated resin of the proper amount of either 2,-diphenyloxazole (PPO) or 2-(4-tert-Butylphenyl)-5-(4-phenylphenyl)-1,3,4-oxadiazole (butyl-PBD) as primary dyes, whereas Lumogen Violet (LV, BASF) has been chosen as waveshifting dye. For selected formulations the addition of o-carborane ($C_2B_{10}H_{12}$, Katchem Ltd, used as received) was also performed, as described in Table I. The obtained cylindrical samples appeared as perfectly transparent and mechanically stable and firm. According to the details given in Table I, these samples have been labeled as A-cond18, while labels A22 and A100 are referred to samples obtained from the traditional addition Pt-catalyzed reaction, as described in previous papers [7-9,12].

PPO and butyl-PBD displayed optimal solubility in the aromatic siloxane system, owing to the presence of phenyl substituents along the siloxane chain, which increase chemical affinity with the aromatic structure of both dyes, as can be deduced by their chemical structures (not reported). Concentrations as high as 6% wt. of PPO have been reached without evident symptoms of precipitation, as reported in Table I. Good dyes dispersion inside the base matrix is a key requirement for the achievement of an organic scintillator since aggregation invariably induces light scattering and yield loss. On the other hand, the good compatibility between the emission range of the primary dye and the excitation maximum of the waveshifter is also crucial in order to shift the final emission in the wavelength range of maximum sensitivity of photo-sensors. In this case, LV absorbs where PPO and butyl-PBD display the maximum emission (320-360 nm) in the UV and re-emits in the violet-blue region (about 430 nm), where the maximum sensitivity of standard PMTs falls. On the other hand, very recently encouraging results have been presented as related to coupling of red-emitting polysiloxane scintillators with less costly and more durable avalanche photo diodes [11].

Table I. Composition of the PSS obtained from addition and condensation reaction and tested as scintillators in this work.

| Resin (A) | Dyes (% wt.) | | Resin B | Catalyst |
|---|---|---|---|---|
| A100 | LV 0.02% | PPO: 1%, 2%, 3%, 4%, 5%, 6% <br> b-PBD: 0.1%, 0.2%, 0.3%, 0.5%, 1% | methylhydro siloxane hydride terminated | Pt 1.2 µL/g |
| A22 | LV 0.02% | PPO: 1%,2%,3% <br> b-PBD: 0.1%,0.2%,0.3%,0.4%,0.5%,1% | methylhydro siloxane hydride terminated | Pt 1 µL/g |
| Acond18 | LV 0.02% | PPO: 1%, 2%, 3%, 4%, 5%, 6% <br> b-PBD: 0.1%,0.3%,0.5% | Poly-diethoxysilane(PDS) | DBTL 1.3% wt. |
| Acond 18 | LV 0.02% | PPO: 1% <br> o-carborane: 4%, 5%, 6% | Poly-diethoxysilane(PDS) | DBTL 1.3% wt. |

Scintillation measurements were performed by exciting the samples with a $^{241}$Am α source (3 kBq, 5.484 MeV) and a $^{60}$Co γ-ray source. Pulse height spectra were obtained by coupling the scintillator samples to an H6524 Hamamatsu PMT. The sample yields were compared with that one obtained from a reference EJ-212 plastic scintillator (Eljen Technology Products) in the same experimental conditions [7], whereas for B doped samples the yield was also compared with EJ-254, which contains 5%wt. of natural boron.

High Resolution X-Ray Diffraction (HRXRD) patterns were recorded using a Philips X'Pert PRO MRD diffractometer equipped with a Cu Kα1 radiation (~8 keV) as the probe.

## Results and discussion

In Table II the light yield obtained as described in the experimental section from the newly developed scintillators, cross-linked by condensation, is reported. The value is expressed as light yield percentage with respect to the yield measured from either the standard EJ212 or EJ254. The comparative yields demonstrate that, in the case of condensation samples, very good results can be obtained, with relative light output up to 90% of commercial plastic scintillators, in the case of $^{nat}$B doped samples, even when the doping percentage is higher in our samples (6% with respect to 5%). Some previous results under neutron irradiation have been reported in [12].

The results after irradiation with the $^{241}$Am α-source (top) and with the $^{60}$Co γ-source (bottom) are shown in Fig. 3 (top panel) as a function of growing percentage of b-PBD. An increase of the light output up to 0.5% of b-PBD can be observed, reaching a value similar to that one obtained with the sample at the limit of saturation: over this limit b-PBD evidences lack of solubility. A general better result can be observed for the A18cond samples with respect to the samples made by addition: this is more easily noticed for the 0.3% percentage case.

Table II. Light output (L.O.) for the newly developed condensation scintillators, compared to standard EJ212 or EJ254 scintillators.

| Base Resin<br>A:B:Sncat (wt. ratio) | Acond<br>100:20:1.3 | Acond<br>100:20:1.3 | Acond<br>100:20:1.3 | Acond<br>100:20:1.3 |
|---|---|---|---|---|
| PPO (% wt.) | 1 | 1 | 1 | 1 |
| LV (% wt.) | 0.02 | 0.02 | 0.02 | 0.02 |
| $^{nat}$B (% wt.) |  | 4% | 5% | 6% |
| $^{241}$Am |  |  |  |  |
| %L.O vs EJ212 | 56±16 | 49±15 | 42±13 | 48±15 |
| %L.O vs EJ254 |  | 73±25 | 63±22 | 71±24 |
| $^{60}$Co |  |  |  |  |
| %L.O vs EJ212 | 68±15 | 70±17 | 61±15 | 68±15 |
| %L.O vs EJ254 |  | 90±22 | 79±20 | 87±19 |

This may be explained with the higher transparency of the samples prepared by condensation, while a light yellow appearance can be seen in the addition samples, probably due to platinum nanoclusters formation as explained previously.

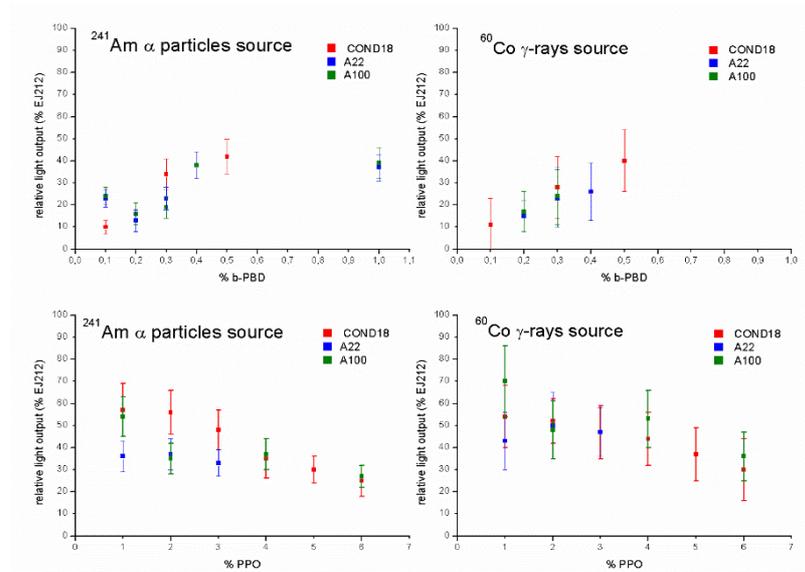

Fig. 3. Light yield under α and γ irradiation of samples added with b-PBD (top panel) and PPO (bottom panel): A18Cond, A100, A22. See Table I and text for more information.

In Fig. 3 (bottom panel) the light output yield is shown for the second series of samples as a function of growing PPO percentage. The percentage of primary dye is in this case much larger than in the case of b-PBD, owing to higher solubility.

Samples containing very high PPO concentration (up to 6%) still display good light yield under exposure to α-particle and γ-rays irradiation (~40% vs. EJ212). It has to be noticed that only up to 3% of PPO can be dispersed in the A22 resin, while in A100/A18cond the solubility is still good up to 6% in weight. PPO showed solubility problems in PSS over 10% wt. as white residue can be observed mainly on the surface due to the diffusion of saturated PPO.

The results on the Light Output obtained with these samples have similar values of those obtained with samples made two years ago and recently re-measured. On one side reproducible results can be obtained following the synthesis procedure, on the other side negligible ageing effects are observed even at high levels of PPO concentrations, which are useful for PSD [13].

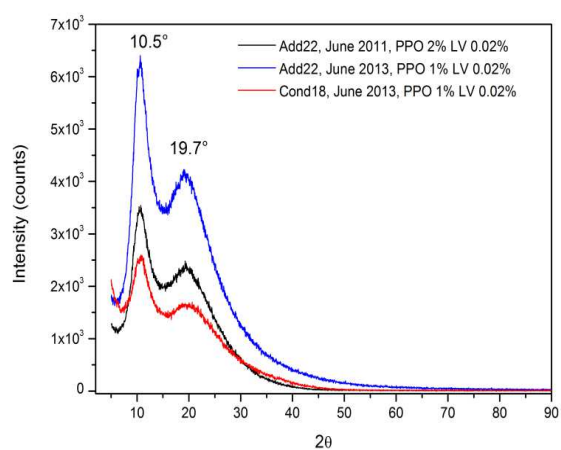

Fig. 4. X–Ray Diffraction spectra of samples, which differ for type and date of synthesis.

The structure of selected samples has been also investigated by X-ray diffraction: samples A22 were prepared in different periods (June 2011 & June 2013) whereas Cond18 were synthesized by condensation (June 2013). No sharp peaks are evident, indicating the absence of 3D crystalline order in all the samples, as expected. Among the three samples, only negligible changes are evident with the different polymerization routes. Two components are observed: the peak at 19.7° is related to the Si-O amorphous phase of silica-like structure in siloxane, the peak at 10.7° is related to an ordered domain (lamellae with folded chains) without a close crystalline packing.

## Conclusions

In this paper the main results with polysiloxane based scintillators are summarized in order to evidence the good light response under α-particle and γ-rays irradiation as compared to standard plastics. Different primary dyes have been added to both addition and condensation cross-linked siloxanes showing in all cases good solubility, mechanical robustness and overall scintillating performances. In particular, systems based on polycondensation proved to be highly promising as for optical transparency, dye dispersion capability and long-lasting light output. Very high PPO concentrations have been achieved within this systems preserving all the mentioned features: PSD capabilities of these scintillators are actually being tested with both fast and slow neutrons.